# Microcracks and inhomogeneously distributed defects in solids

**Yuri Kornyushin**

*Maitre Jean Brunschvig Research Unit*
*Chalet Shalva, Randogne, 3975-CH*

**Abstract**

A conception of inhomogeneous locally random distribution of microdefects in crystalline solids is proposed. A method to calculate some physical properties of solids, containing inhomogeneously distributed defects, is developed. A contribution of this inhomogeneity to a series of physical properties is calculated and discussed. This contribution exceeds that of homogeneously distributed defects by the orders of magnitude. A contribution of the inhomogeneity to electric conductivity, Hall effect and magnetoresistance is calculated. Elastic energy and volume of inhomogeneously dislocated crystal are regarded. It was shown that the relaxation of the elastic energy of random dislocations during propagation of a crack facilitates the process. These results explained a phenomenon of lamination of overdeformed metals.

## 1. Introduction

A conception of inhomogeneous locally random distribution of defects was proposed by the author in 1966 [1,2] and its application to calculation of some physical properties is given in [3]. This conception proved to be a rather fruitful one, it allowed calculating in details effective electric conductivity, Hall effect coefficient and magetoresistance of metals, containing inhomogeneously distributed dislocations, and semiconductors (including multi-valley ones), containing inhomogeneously distributed charged point defects and charged dislocations [3]. In this paper the author will skip some basic features, concerning galvanomagnetic properties, and concentrate on elastic energy of dislocated crystal and fracture [5].

The value of the stored elastic energy is very important for the mechanical properties of solid materials and fracture processes occurring in them. One aim of the work was to develop a general method to calculate the elastic energy of a crystal containing arbitrarily distributed but locally random dislocations.

Various types of inhomogeneous distribution of dislocations are observed in highly deformed crystalline materials, in particular, pile-ups, cellular structure, etc. To develop a general method of calculation of the energy of such systems a plane wave Fourier expansion technique was used. This approach was used earlier in the works of A.D. Brailsford, [5]. The technique used in [5] allows for the fulfillment of all the calculations of the elastic energy in general, and to obtain rather clear and simple formulas which could be easily applied to particular cases.

Another advantage of the method developed in [5] is natural and inherent to the solid state physics description of the deformations around a dislocation in the vicinity of its axis. In the theory of elasticity [6] a cut-off radius of a dislocation is usually introduced and this procedure does not look very natural from the point of view of the solid state physics. Using the Fourier technique, A.D. Brailsford introduced a limiting Debye wave vector, $q_D$, instead of a cut-off radius. Both procedures are equivalent, but the latter one looks more logical in solid state theory.

In this paper the inhomogeneous locally random distribution will be defined, the technique of calculation of the elastic energy will be described, propagation of a crack both along the grain boundary and through the bulk of a grain will be considered, special emphasis on the physical explanation of the phenomena of lamination will be made, and application of the results of testing of highly deformed crystals in respect to the presence of the microcracks will be given.

**2. Inhomogeneous Locally Random Distribution**

Let us consider a model of distribution of linear dislocations which takes into account both random fluctuations of the dislocation density around its average value, $\langle n \rangle$, as well as its regular dependence on coordinates, $n(\mathbf{r}_s)$ (here $\mathbf{r}_s$ describes the location of the site s) [3]. This model is a generalization of the random distribution for two-dimensional problems. It is based on the following assumptions:
1) Linear dislocation has a definite position within the plane elementary cell of a crystal.
2) Two different parallel dislocations cannot take the same position.
3) Two different dislocations which are not parallel could cross freely.
4) The probability of the event that a given dislocation of a definite type occupies $\mathbf{r}_s$ position is

$$p(\mathbf{r}_s) = sn(\mathbf{r}_s)/S\langle n \rangle, \qquad (1)$$

where $s$ is the area of the plane elementary cell, $S$ is the area of the section of a sample (grain in a polycrystal), perpendicular to the dislocation line.

Eq. (1) represents a generalization of the random distribution. When $n(\mathbf{r}_s) \equiv \langle n \rangle$ Eq. (1) describes a regular random distribution of dislocations over all the section of a sample. So really Eq. (1) describes a case when usual random distribution is modulated by a coordinate-dependent factor. The section of a sample could be divided into small parts; in each one a coordinate dependence of the probability is inessential. So the regarded distribution may be called a locally random one. Modulating function is determined by the history of the preparation of a sample and determines a spatial mesoscopic inhomogeneity of it.

**3. Calculation of the Elastic Energy**

Elastic energy, $W$, is given by the integration over the volume of a crystal (grain in a polycrystal), $V$, the density of the elastic energy, $w$, which is represented by the sum over the indices, i, k, l, m, from 1 to 3 of the expression [6]

$$w(\mathbf{r}) = 0.5 c_{iklm} u_{ik}(\mathbf{r}) u_{lm}(\mathbf{r}), \qquad (2)$$

where $c_{iklm}$ is the elastic moduli and $u_{ik}$ is the tensor of deformations.

Boundary conditions do not contribute essentially into elastic fields of dislocations [7]. Neglecting this contribution the Fourier components of the deformations belonging to the dislocation s may be represented as

$$u_{s,ik\mathbf{q}} = u_{ik\mathbf{q}} \exp{-i(\mathbf{q},\mathbf{r}_s)}, \qquad (3)$$

where $u_{ik\mathbf{q}}$ is a Fourier component of deformations belonging to a linear dislocation situated in zero coordinates; magnitude of the Fourier vector, $\mathbf{q}$, varies from $2\pi/d$ to $q_D$ ($d$ is the size of a crystal or grain size in a polycrystal).

Using Eq. (3) one may see that the contribution of the spatial distribution of dislocations to elastic energy is described by the structural factor

$$F = \sum_{sp} \exp i(\mathbf{q}, \mathbf{r}_s - \mathbf{r}_p), \tag{4}$$

which may be calculated by the way of averaging it over $\mathbf{r}_s$ and $\mathbf{r}_p$ using Eq. (1) for smooth functions $n(\mathbf{r})$, when $b|\nabla n| \ll n$ ($b$ is the value of the Burgers vector).

In the case when s and p refer to the dislocations of one type

$$F = d^2 \langle n \rangle + d^4 n_\mathbf{q} n_{-\mathbf{q}}. \tag{5}$$

More detailed consideration of this and other cases is given in [1,3]. The first term in Eq. (5), $d^2\langle n \rangle$, represents a contribution of each separate dislocation. The second term, $d^4 n_\mathbf{q} n_{-\mathbf{q}}$, describes the contribution of the smooth additional field of deformation arising due to a spatial inhomogeneous distribution of dislocations and represents an interaction between dislocations. In case of random distribution the second term in Eq. (5) vanishes and elastic energy is expressed by the sum of elastic energies of separate dislocations, each term having a factor $\ln(dq_D/2\pi)$ ($2\pi/q_D$ is equivalent to the cut-off radius), which corresponds to the absence of interaction between stress fields of separate random dislocations. This means that the stress fields of random dislocations are the long-range ones.

The stress fields of the ordered dislocations, on the contrary, are the short-range ones [6]. Accordingly, the expression for the elastic energy of the ordered dislocations has a factor $\ln(hq_D/2\pi)$ [6], instead of $\ln(dq_D/2\pi)$ ($h \ll d$ is the spacing between the ordered dislocations). That is why the elastic energy of the random homogeneously distributed dislocations is several times larger than that of the ordered ones.

To estimate the contribution of inhomogeneity to elastic energy let us consider an example of the periodical distribution with the period, $l$:

$$n(x) = \langle n \rangle[1 + a\cos(2\pi x/l)], \quad 0 < a < 1. \tag{6}$$

In this case a relative contribution of inhomogeneity is as follows:

$$f = 4\langle n \rangle (al)^2/3\pi\ln(dq_D/2\pi). \tag{7}$$

At $l = 5\mu$, $a = 0.1$, $\langle n \rangle = 10^{11}$ cm$^{-2}$, $\ln(dq_D/2\pi) = 5$, Eq. (5) yields $f = 20$. So the contribution of inhomogeneity in this case is very essential. More detailed information is published in [1,3].

## 4. Relaxation of the Elastic Energy during Propagation of a Crack

When crack propagates, the area of the free surface increases. As on the surface of a crack we have zero boundary conditions, increase in the area of a surface leads to a partial relaxation of elastic stresses near the free surface and elastic energy of a grain decreases. In case of a crystal, containing random dislocations, stress fields of which are the long-range ones, and reach the surface of a crack, deformations produced by each random dislocation relax. This relaxation contributes to the energetic balance, which controls a propagation of a crack, and facilitates the propagation.

The description of the process is different in cases of fine-grained material (where the size of a crack exceeds essentially the grain size) and coarse-grained material (where the development of a crack occurs in the limits of a single grain). These two cases ought to be regarded separately.

4.1. Fine-Grained Material

As was mentioned above, elastic energy of random dislocations in a grain consists of the energies of separate dislocations calculated as if each dislocation is the only one in a grain and is situated in the central part of it. During propagation of a crack, when the free surface area increases, stress field of dislocations, situated in a grain where a free surface arises, partly relax. This leads to a decrease in the effective specific surface energy, $\alpha$ [8]. Stress relaxation in grains not neighboring fracture surface is insignificant [8] and will not be taken into account. There are two cases:

4.1.1. *Intergranular Propagation.* In this case the effective decrease in the specific surface energy is caused only by new free boundary conditions on the fracture surface and is described by the formula [8]:

$$\alpha = \alpha_0 - kndGb^2/24\pi, \tag{8}$$

where $\alpha$ is the effective specific surface energy in a dislocated crystal, $\alpha_0$ is the specific surface energy in a perfect crystal, $k$ is some factor of the order of unity (details may be found in [6,8]), and $G$ is the shear modulus.

At $k = 1$, $G = 160$ GPa, $b = 3\times10^{-8}$ cm, $n = 10^{10}$ cm$^{-2}$, $d = 50$ μm we have $\alpha - \alpha_0 = 1.06$ N/m. So the contribution of the random dislocations is very essential.

In case of ordered dislocations their stress fields is not a long-range ones and only the stress fields of the nearest to the surface dislocations relax during the increase of the surface area. That is why the contribution of the ordered dislocations, which have a short-range stress fields, to the effective specific surface energy is negligible.

4.1.2. *Transgranular Propagation.* In this case the decrease in the effective specific surface energy is caused by the new boundary conditions on the surface of a crack as well as by the decrease in the grain size (at least in one direction). The effective decrease in the specific surface energy may be evaluated by the formula [8]

$$\alpha = \alpha_0 - kndGb^2(1 + m)/24\pi, \tag{9}$$

where $m$ is some constant, the value of which is somewhere between 2 and 4.

The right-hand part of Eq. (9) has an additional factor $(1 + m) > 1$ as compared to Eq. (8). So the transgranular propagation is facilitated by the random dislocations more essentially than the intergranular one. But the preference of the type of propagation is determined not only by the value of the change in the effective specific surface energy, $\alpha - \alpha_0$, but also by the value $\alpha_0$ itself and other factors.

4.2. Coarse-Grained Material

When development of a crack occurs in the limits of a single grain, the contribution of the relaxation of the elastic energy of the random dislocations is also essential, but it may not be described by the terms of the effective specific surface energy [9]. A major part of the elastic energy of a given random dislocation is concentrated within the limits of a grain where the dislocation is situated [9]. Let us consider two cases again:

4.2.1. *Intergranular Propagation.* When a crack opens along a grain boundary, the surface of a crack becomes a free surface, and so the boundary conditions on this area of the grain change from constraint to free ones. This leads to the partial relaxation of the elastic energy of random dislocations. This relaxation may be evaluated using St. Venant principle [6]. The contribution of randomly distributed immobile dislocations to the relaxation of the elastic energy of a crystal may be accounted for by introduction of the effective tensile stress, $P$, according to the formula [9]

$$P^2 = P_e^2 + AG^2b^2n, \tag{10}$$

where $P_e$ is the external tensile stress, and factor $A$ for the case of a grain boundary crack is as follows

$$A_i = 2k((1/6) + \{[\ln(d/L)]/[\ln(dq_D/2\pi)]\})/\pi^2(1 - \sigma), \tag{11}$$

where $L$ is a linear size (height) of a crack, and $\sigma$ is the Poisson' ratio.

Typical value of the factor $A_i$ is 0.01 and the contribution of the dislocations to the critical size of a crack becomes essential when $n \geq 10^9$ cm$^{-2}$.

4.2.2. *Transgranular Propagation.* When a crack opens inside a grain, free surface arises there. Elastic energy of a given random dislocation situated at a distance $x < L$ from the surface of a crack, contains a factor $\ln(xq_D/2\pi)$ instead of the factor $\ln(dq_D/2\pi)$ for the dislocation situated in a crack free grain. As $x < d$, the elastic energy of a dislocation decreases. Stress field of dislocation situated at a distance $x > L$ relaxes in the area of a linear size $L$ around a crack. Again, using St. Venant principle, it is possible to evaluate this relaxation and obtain again Eq. (10) with

$$A = A_t = 5km[\ln(d/L)]/4\pi^2(1 - \sigma). \tag{12}$$

Typical value of the factor $A_t$ according to Eq. (12) is several times (up to the order of magnitude) larger than that of $A_i$ according to Eq. (11). This means greater influence of the random dislocations upon development of the cracks inside a grain compared to the case of grain boundary cracks.

**5. Lamination**

Dislocations in over-rolled refractory metals and alloys usually form a cellular structure. Fracture of material often occurs along the cell boundaries. Let us consider a cellular dislocation structure. Let us assume that the cell boundaries are layers or walls, consisting of random dislocations. It is well known that the elastic energy of the ordered dislocations released during fracture is insufficient to provide a spontaneous fracture of a material. Elastic energy of dislocations randomly situated in the layer or wall is represented by the expression [9]

$$W = kmGdb^2N\{[\pi(\delta N)^2/N] + [\ln(dq_D/2\pi)]\}/8\pi, \tag{13}$$

where $N$ is the total number of the random dislocations in the layer or the wall, $\delta N$ is the surplus of the dislocations of one sign. The relaxation of the elastic energy [Eq. (13)] can be and often is larger than that, which is needed to compensate the increase in the surface energy during the process of lamination. On the contrary, when dislocations in the layer (wall) are equidistantly situated, there is only a second term in the brackets in the right-hand part of Eq. (13), and instead of the grain size, $d$, there is a spacing between the equidistant dislocations, $h \ll d$ [6]. Accordingly, relaxation of the elastic energy cannot compensate increase in the surface energy at the values of parameters ($N$, $G$, $b$, etc.), which may be practically realized. From this follows that under such circumstances lamination is not possible. Indeed, for the regular values of parameters the relaxation of the elastic energy can hardly compensate 0.2 N/m of the specific surface energy, because typical value of it is 1 - 2 N/m (5 - 10 times larger). For the random dislocations even at $\delta N = 0$ and at typical value parameters, the value of the released stored energy is high enough to compensate surface energy of the surface area growing during the process of lamination.

## 6. Wedge Microcrack

Wedge cracks formed from edge dislocations are the basic carrier of the fracture in crystalline materials. It is well know that the thermodynamics of cracks and dislocations is determined by the energy balance, since entropy terms are negligible [6]. The existence of a free boundary in the crack cavity produces a relaxation of the elastic energy resulting from external loading. The contribution of the randomly distributed immobile dislocations in the relaxation of the elastic energy of a system has been considered above. It may be described either in terms of the effective specific surface energy (for fine-grained materials), or by introducing effective tensile stress (for coarse-grained materials). Changes in the surface area and corresponding surface energy at variation of the crack length ought to be taken into account also. Details of the thermodynamic analysis of the development of a crack are described in [10].

### 6.1. Equilibrium Height of a Crack

The equilibrium height of a crack, $L$, is determined by the condition of a minimum of the Gibbs free energy and is given by the formula

$$L = [4G\alpha/\pi(1 - \sigma)P^2]\{1 - [1 - (Pnb/4\alpha)^2]^{1/2}\}, \qquad (14)$$

where $N$ is the power of a crack.

In the case of an extremely low $P$ Eq. (14) results in the well-known formula for the height of the wedge crack, assuming that neither load nor dislocations are present.

### 6.2. Critical Parameter of a Crack

The unsteady state of a crack occurs when the second derivative of the Gibbs free energy with respect to the crack height becomes equal to zero. This corresponds to the critical value of the effective stress,

$$P_c = 4\alpha/Nb. \qquad (15)$$

It is worthwhile to note that the critical crack height, corresponding to Eq. (15), is equal to the half of the Griffith critical crack size (with regard to the randomly distributed immobile dislocations [9]).

### 6.3. Stress-Elongation Relationship

As the load grows, the wedge crack reversibly elongates, doubling its zero load height by the moment it becomes unstable. This contributes to the elongation of the sample and leads to the reversible elastic sublinearity caused by the growth in the size of brittle steady cracks during loading. At a microcrack density $3 \times 10^5$ cm$^{-2}$, $N = 60$ and other typical values of parameters deviation from the Hooke's law near the critical load consists of about 7%. Such sublinear stress-elongation relationship may be observed in cracked ceramic materials.

## 7. Tests on Microcracks

There is a limited value of plastic deformation of a crystalline material, after which a material still remains free of cracks. The limiting value corresponds to the beginning of the formation of microcracks, which leads to decrease in the strength of a material. The method to determine the plasticity limit of a metal under mechanical treatment was proposed in [11]. This method is based upon comparison of changes in the stored energy and in the density of a material occurring during mechanical treatment. Density changes due to dislocations are determined by the terms quadratic on the components of the deformation tensor (analogous to that for the stored elastic energy [12]). Hence the ratio of the stored energy to the density change varies only slightly (less than about 10%) for various possible distributions of dislocations. Formation of cracks leads to the sharp increase in the volume change due to the volume of the crack cavities, whereas volume change due to the deformations around cracks remain proportional to their elastic energy, with practically the same coefficient as for the random or equidistant dislocations. Contribution of the surface energy of the cavities of the cracks into the total energy remains negligible. So the energy to volume ratio undergoes a distinguishable deviation from the constant value when microcracks appear in a material.

Typical value of the changes in the volume due to the elastic deformations in a dislocated crystal is about $10^{-4}$ of the volume of a sample. Additional change of about 20% from this value is enough to recognize a starting process of the formation of cracks.

So the proposed method allows one to determine experimentally the appearance of the cavities, whose volume constitutes of about $2 \times 10^{-5}$ of the volume of a sample.

## 8. Some Other Physical Properties

As was mentioned above, the dependence of the changes in the volume of a crystal on the components of the deformation tensor related to dislocations is of the same nature as that for the stored elastic energy [12]. From this follows that all the discussed peculiarities of the elastic energy of various dislocation structures refer also to the changes in the volume of a crystal. The same considerations apply also to the contribution of dislocations to the electric conductivity of a crystal.

In the case of charged point defects in semiconductors the problem becomes three-dimensional, and in Eq. (1) $s$ and $S$ should be replaced by $v$ (a volume of elementary cell) and $V$ (a volume of a sample). In this case Eqs. (3 - 5) acquire a three-dimensional meaning and all the calculations become three-dimensional ones [3]. It is worthwhile to note that the second term in Eq. (5) describes the contribution of a spatial electric charge arising due to inhomogeneous distribution of the charged point defects. So the contribution of the charged point defects consists of two parts: the first one represents the sum of the contribution of each point defect (independently on their distribution), and the second one describes the contribution of the mentioned spatial electric charge. Details are available in [3].

It should be mentioned that the problem of the effective electric conductivity of inhomogeneous materials has a long history, which is reported in [3,13]. In particular for the granular-type inhomogeneous materials it is possible to use the Clausius-Mossotti approximation to evaluate the effective conductivity of a sample [13].

There is also a number of exact solutions in the problem of the effective electric conductivity of inhomogeneous samples. These solutions are reported and discussed in [3].

## 9. Discussion

A model of randomly distributed dislocations was considered here and in [16]. Random distribution was generalized for the spatially inhomogeneous case when random distribution is modulated by smooth function of coordinates. This model is relevant for the samples deformed at relatively low temperatures (lower than 0.5 of the melting point). Dislocation ensembles in real samples may be regarded as consisting of equidistant dislocations along with the random ones, i.e. part of the dislocations of the ensemble may be regarded as random ones, and another part - as an ordered dislocations. Random dislocations distort the X-ray scattering picture of a crystal in a special way. Experimental data on X-ray scattering and others, discussed in [8] confirm the presence of the random dislocations in cold-deformed materials, in particular, in the cell boundaries of over-rolled refractory metals and alloys. So the described results allowed for a self-consistent explanation of various experiments performed on dislocated crystals, in particular, they allowed the development of the method to determine the limit of plasticity of metallic materials which was used to improve mechanical treatment of various metals and alloys [14,15].

The described sublinear stress-strain relation due to the reversible elongation of cracks may be useful in ceramic materials, making possible a partial relaxation of stresses in overloaded areas.

The electric conductivity measurements of a high accuracy are used as one of the methods of investigation of the defect structure of materials.